\documentclass[aps,prd,twocolumn,showpacs,preprintnumbers,amsmath,amssymb]{revtex4}

\usepackage{graphics,color}
\usepackage{epsfig}
\usepackage{dcolumn}
\usepackage{bm}

\begin{document}
\title{Is $X(3872)$ {\sl Really} a Molecular State?}

\author{Yan-Rui Liu$^{1}$}
\email{yrliu@ihep.ac.cn}

\author{Xiang Liu$^2$}
\email{xiangliu@pku.edu.cn}

\author{Wei-Zhen Deng$^2$}
\email{dwz@th.phy.pku.edu.cn}

\author{Shi-Lin Zhu$^2$\footnote{Corresponding author}}
\email{zhusl@phy.pku.edu.cn}

\affiliation{$^1$Institute of High Energy Physics, P.O. Box 918-4,
Beijing 100049, China\\ $^2$Department of Physics, Peking
University, Beijing 100871, China}

\date{\today}

\begin{abstract}

After taking into account both the pion and sigma meson exchange
potential, we have performed a dynamical calculation of the
$D^0\bar{D}^{\ast0}$ system. The $\sigma$ meson exchange potential
is repulsive from heavy quark symmetry and numerically important for
a loosely bound system. Our analysis disfavors the interpretation of
X(3872) as a loosely bound molecular state if we use the
experimental $D^\ast D\pi$ coupling constant $g=0.59$ and a
reasonable cutoff around 1 GeV, which is the typical hadronic scale.
Bound state solutions with negative eigenvalues for the
$D\bar{D}^\ast$ system exist only with either a very large coupling
constant (two times of the experimental value) or a large cutoff
($\Lambda \sim 6$ GeV or $\beta \sim 6$ GeV$^2$). In contrast, there
probably exists a loosely bound S-wave $B\bar{B}^\ast$ molecular
state. Once produced, such a molecular state would be rather stable
since its dominant decay mode is the radiative decay through
$B^\ast\to B \gamma$. Experimental search of these states will be
very interesting.

\end{abstract}


\maketitle

\section{introduction}\label{sec1}

Since the observation of the charmonium-like state $X(3872)$ in
the $J/\psi\pi^+\pi^-$ channel by Belle collaboration in 2003
\cite{3872-first}, $X(3872)$ has been confirmed by CDF
\cite{3872-CDF}, D0 \cite{3872-D0} and Babar collaborations
\cite{3872-Babar}. In the past three years, there have accumulated
abundant experimental information of $X(3872)$, which is collected
in Table \ref{3872-review}.
\begin{table}[htb]
\begin{tabular}{|c||c|}
\hline& $X(3872)$
\\\hline\hline
& $3872.0\pm0.6\pm0.5$ \cite{3872-first}\\
&$3871.3\pm0.7\pm0.4$ \cite{3872-CDF}\\
Mass&$3871.8\pm 3.1\pm 3.0$ \cite{3872-D0}\\
(MeV)&$3873.4\pm1.4$ \cite{3872-Babar}\\
&$3875.4\pm0.7^{+1.2}_{-2.0}$ \cite{3872-Belle-3875}\\
&$3875.6\pm0.7^{+1.4}_{-1.5}$ \cite{3872-Babar-3875}\\
\hline Width&$<2.3$ MeV \cite{3872-first}  \\
\hline$ J^{PC} $&$1^{++}/2^{-+}$ \cite{3872-angular,3872-angular-CDF}   \\
\hline
&$X(3872)\to J/\psi \pi^+\pi^-$ \cite{3872-first,3872-CDF,3872-D0,3872-Babar} \\
&$X(3872)\to \gamma J/\psi,\omega J/\psi$ \cite{3872-gamma,3872-gamma-Babar}\\
Decay channels&$X(3872)\to \rho J/\psi$ \cite{3872-rho-CDF}\\
&$X(3872)\to D^0 \bar{D}^0 \pi^0$ \cite{3872-Belle-3875}\\
&$X(3872)\to D^0 \bar{D}^{*0}+h.c.$ \cite{3872-Babar-3875}\\
\hline &$\frac{BR[X(3872)\to \gamma J/\psi]}{BR[X(3872)\to \pi^+\pi^- J/\psi]}=0.14\pm0.05$ \cite{3872-gamma} \\
Branching fractions&$\frac{BR[X(3872)\to \gamma J/\psi]}{BR[X(3872)\to \pi^+\pi^- J/\psi]}=0.25$ \cite{3872-gamma-Babar}\\
&$\frac{BR[X(3872)\to D^0\bar{D}^0\pi^0]}{BR[X(3872)\to \pi^+\pi^-
J/\psi]}=9.4^{+3.6}_{-4.3}$ \cite{3872-Belle-3875}
\\\hline
\end{tabular}
\caption{A review of the experimental status of $X(3872)$.
\label{3872-review}}
\end{table}

Quark model calculation indicates that a $2^3P_1$ $c\bar{c}$ state
$\chi_{c1}'$ lies $50\sim 200$ MeV above $X(3872)$. Moreover a
charmonium state with isospin $I=0$ does not decay into $J/\psi\rho$
easily. Thus there is some difficulty of the charmonium assignment
of $X(3872)$. The possible theoretical explanations of $X(3872)$
include a molecule state
\cite{3872-Mole-1,3872-Mole-2,3872-Mole-3,3872-Mole-4,3872-Mole-5},
a $1^{++}$ cusp \cite{3872-cusp}, the S-wave threshold effect due to
the $D^0\bar{D}^{0*}$ threshold \cite{3872-s-wave}, a hybrid
charmonium \cite{3872-hybrid}, a diquark anti-diquark bound state
\cite{3872-D}, a tetraquark state \cite{3872-tetra} and a
dynamically generated resonance \cite{oset}.

Among these theoretical schemes, the molecule picture is the most
popular one due to the following reasons. The molecular picture
naturally explains both the proximity of $X(3872)$ to the
$D^0\bar{D}^{*0}$ threshold and the isospin violating $J/\psi \rho$
decay mode. It predicted the decay width of the
$J/\psi\pi^+\pi^-\pi^0$ mode to be comparable with that of
$J/\psi\rho$, which was confirmed by Belle collaboration
\cite{3872-gamma}. Within the same picture, Braaten and Kusunoki
predicted that the branching ratio of $B^0\to X(3872)K^0$ is
suppressed by more than one order of magnitude compared to that of
$B^+\to X(3872)K^+$ \cite{brateen 3872}.

Later both Belle and Babar collaborations observed the radiative
decay mode. Belle's measurement found \cite{3872-gamma}
\begin{equation}
\frac{BR[X(3872)\to \gamma J/\psi]}{BR[X(3872)\to J/\psi
\pi^+\pi^-]}=0.14\pm 0.05
\end{equation}
while Babar collaboration got \cite{3872-gamma-Babar}
\begin{equation}
\frac{BR[X(3872)\to \gamma J/\psi]}{BR[X(3872)\to J/\psi
\pi^+\pi^-]}\approx0.25 \; ,
\end{equation}
which are against the prediction from the molecular picture
$7\times 10^{-3}$.

Recently Belle collaboration measured the ratio
\cite{3872-Belle-3875}
\begin{equation}
\frac{BR[X(3872)\to D^0\bar{D}^0\pi^0]}{BR[X(3872)\to \pi^+\pi^-
J/\psi]}=9.4^{+3.6}_{-4.3}
\end{equation}
which is much larger than the theoretical value $0.054$ from the
molecular assumption. From Ref. \cite{3872-Belle-3875}, one can
also extract
\begin{equation}
\frac{BR[B^0\to X(3872)K^0]}{BR[B^+\to X(3872)K^+]}\approx 1.62
\end{equation}
which is also much larger than the molecule prediction.

Up to now, several groups carried out the dynamical study of the
molecular assignment of $X(3872)$. Swanson proposed that $X(3872)$
was mainly a $D^0 \bar{D}^{*0}$ molecule bound by both the pion
exchange and quark exchange \cite{3872-Mole-4}. To obtain the
potential between $D^0 \bar{D}^{*0}$ through exchanging single
pion, he followed the method proposed by T\"{o}rnqvist
\cite{Tornqvist}. The formalism is based on a microscopic
quark-pion interaction. Swanson indicated that one pion exchange
alone can not bind $D$ and $D^*$. He also included the short-range
quark-gluon force \cite{3872-Mole-4}.

In Ref. \cite{3872-Mole-3}, Wong studied the $DD^*$ system in the
quark model in terms of a four-body non-relativistic Hamiltonian
with pairwise effective interactions. This framework is similar to
the consideration of adding short-range quark-gluon force in
Swanson's paper \cite{3872-Mole-4}. The author found an S-wave
$DD^*$ molecule with the binding energy $\sim 7.53$ MeV. In Refs.
\cite{MAGP,FKMK,BML,HKKN,V,cola}, further investigations basing on
the molecular assumption are carried out.

With the obtained one pion exchange potential (OPEP) by using the
effective Lagrangian, Suzuki argued that $X(3872)$ is not a
molecular state of $D^0\bar{D}^{*0}+\bar{D}^0 D^{*0}$ \cite{suzuki},
which contradicts Swanson and Wong's conclusion. Instead, $X(3872)$
may have a dominant $c\bar{c}$ component with some admixture of
$D^0\bar{D}^{*0}+\bar{D}^0 D^{*0}$
\cite{chao-3872,suzuki,zhu-review}.

In order to further clarify the underlying structure of X(3872),
we shall carry out a systematic dynamical study of the molecular
picture in this work. It's important to note that the one pion
exchange potential alone does not bind the proton and neutron pair
into the deuteron in nuclear physics. In fact, the strong
attractive force in the intermediate range has to be introduced in
order to bind the deuteron, which is modelled by the sigma meson
exchange potential elegantly. We shall explore whether the similar
mechanism plays an important role in the case of $X(3872)$.

This work is organized as follows. After the introduction, we give a
concise review of the molecular picture. In Section \ref{sec2} we
present the flavor wave function of X(3772), effective Lagrangian
and coupling constants relevant to the derivation of the $\pi$ and
$\sigma$ exchange potentials. In Section \ref{sec3}, we illustrate
the procedure to obtain the potentials and give their expressions.
Then we present the numerical results in Section \ref{sec4} and
\ref{both}. The last section is the summary and discussion.

\section{Review of Molecular Picture}\label{1.1}

In the study of hadron spectroscopy, some states are difficult to
be accommodated in the conventional $q\bar{q}$ and $qqq$
framework. These states are considered good candidates of hadrons
beyond the conventional valence quark model. The possible
assignments include the glueball, hybrid state and mutiquark state
etc. Among them, the molecular state is very attractive.

In the past thirty years, theorists have been studying whether two
charmed mesons can be bound into the molecular state because the
presence of the heavy quarks lowers the kinetic energy while the
interaction between two light quarks could still provide strong
enough attraction. Voloshin and Okun studied the interaction between
a pair of charmed mesons and proposed the possibilities of the
molecular states involving charmed quarks \cite{Okun}. de Rujula,
Georgi and Glashow once suggested $\psi(4040)$ as a $D^*\bar{D}^*$
molecular state \cite{RGG}. T\"{o}rnqvist studied the possible
deuteron-like two-meson bound states such as $D\bar{D}^*$ and
$D^*\bar{D}^*$ using the quark-pion interaction model
\cite{Tornqvist}. Dubynskiy and Voloshin indicated that there exists
a possible new resonance at the $D^*\bar{D}^*$ threshold
\cite{voloshin-1,voloshin}. Besides the above systems, Weinstein and
Isgur studied whether the scalar resonances $f_0(980)$ and
$a_0(980)$ are  molecular states composed of a pair of $K\bar K$
mesons \cite{WI}.

In the past several years, the experimental observations of so
many $X$, $Y$ and $Z$ states stimulated the study of exotic states
greatly. For example, $X(3872)$ is proposed to be a good candidate
of the $DD^*$ molecule state by many groups
\cite{3872-Mole-1,3872-Mole-2,3872-Mole-3,3872-Mole-4,3872-Mole-5},
which is also the topic of the present work. Liu, Zeng and Li
suggested Y(4260) as the $\chi_c\rho^0$ molecule assignment and
predicted its possible decays modes \cite{LZL}. Yuan, Wang and Mo
proposed $Y(4260)$ being a $\chi_{c1}\omega$ molecule \cite{YWM}.
The baryonium possibility was also suggested by Qiao
\cite{qiao-3872}.

Recently Belle collaboration observed a charged state $Z^+(4430)$ in
$\psi'\pi$ channel \cite{Belle-4430}. This new enhancement
immediately triggered the molecular speculation. In fact several
groups suggested $Z^+(4430)$ as a $D_1D^*$ molecular state
\cite{rosner,Meng}. In our previous work \cite{xiangliu}, we have
carried out the first dynamical study of $Z^+(4430)$. Later, we
performed a detailed study of this state in the molecular picture
\cite{lldz}. A short review of the current theoretical status of
$Z^+(4430)$
\cite{rosner,Meng,Bugg,cky,maiani,Gershtein,Qiao,qsr-4430,ding,braaten,liuliudeng}
was also given in Ref. \cite{xiangliu}.

\section{Flavor Wave function, Effective Lagrangian and Coupling
Constants}\label{sec2}

In the following, we will study whether X(3872) is a bound state
of the $DD^*$ meson pair. Before deriving the meson exchange
potential, we first briefly discuss the convention of the flavor
wave function of the molecular state X(3872). In the previous
literature
\cite{3872-Mole-1,3872-Mole-2,3872-Mole-3,3872-Mole-4,3872-Mole-5},
it was defined as
\begin{eqnarray}
|X(3872)\rangle=\frac{1}{\sqrt{2}}\Big[|D^0
\bar{D}^{*0}\rangle+c|{D}^{*0}\bar{D}^0 \rangle\Big]\label{XXwave}
\end{eqnarray}
with $c=+1$. However, this definition does not reflect the
positive C-parity of $X(3872)$ naturally \footnote{We thank E.
Braaten, V. M. Voloshin, E. Swanson and M. Suzuki for useful
communications}. According to the same approach in our previous
paper \cite{xiangliu}, we reanalyze the flavor wave function of
$X(3872)$.

The interpolating current of $X(3872)$ corresponding to Eq.
(\ref{XXwave}) in the quantum field theory reads
\begin{eqnarray}
J_{X(3872)}=\frac{1}{\sqrt{2}}(J_1+c J_2)
\end{eqnarray}
with
\begin{eqnarray*}
J_1&=&(\bar{u}^a\gamma_5 c^a)(\bar{c}^b \gamma^{\mu} u^b),\;\;
J_2=(\bar{c}^a\gamma_5 u^a)(\bar{u}^b \gamma^{\mu} c^b),
\end{eqnarray*}
where $a,b$ denotes the color indices. Under the charge conjugate
transformation, one gets
\begin{eqnarray*}
 {\hat C}J_1 {\hat C}^{-1}=-J_2\;\;\; \mathrm{and}\;\;\; {\hat C} J_2 {\hat C}^{-1}=-J_1.
\end{eqnarray*}
We want to emphasize that there exists no arbitrary phase because
the charm and anti-charm quark and the up and anti-up quark appear
simultaneously. Therefore we obtain
\begin{eqnarray*}
{\hat C}J_{X(3872)} {\hat C}^{-1}=\frac{1}{\sqrt{2}}(-J_2-c J_1).
\end{eqnarray*}
Because the charge parity of $X(3872)$ is $+1$, we have $c=-1$. In
other words, the natural definition of the flavor wave function of
X(3872) should be
\begin{eqnarray}\label{Xwave}
|X(3872)\rangle=\frac{1}{\sqrt{2}}\Big[|D^0
\bar{D}^{*0}\rangle-|{D}^{*0}\bar{D}^0 \rangle\Big].
\end{eqnarray}

In this work, we mainly discuss whether the S-wave $D^0$
($\bar{D}^0$) and $\bar{D}^{*0}$ ($D^{*0}$) molecular state can be
formed by exchanging the $\pi$ and $\sigma$ meson. We need the
effective chiral Lagrangian in the chiral and heavy quark dual
limits \cite{falk,casalbuoni}
\begin{eqnarray}
\mathcal{L} &=&ig {\rm Tr}[H_b {A}\!\!\!\slash_{ba}\gamma_5\bar{H}_a
]+g_\sigma{\rm Tr}[H \sigma\overline{H}]\label{aa}
\end{eqnarray}
with
\begin{eqnarray}
H_a&=&\frac{1+\not v}{2 }[P_{a}^{*\mu}\gamma_\mu-P_a \gamma_5]
\end{eqnarray}
and the axial vector field $A_{ab}^{\mu}$ is defined as
\begin{eqnarray*}
A_{ab}^{\mu}=\frac{1}{2}(\xi^{\dag}\partial^{\mu}\xi-\xi\partial^{\mu}\xi^{\dag})_{ab}=
\frac{i}{f_{\pi}}\partial^{\mu}\mathcal{M}_{ab}+\cdots
\end{eqnarray*}
with $\xi=\exp(i\mathcal{M}/f_{\pi})$, $f_\pi=132$ MeV and
\begin{eqnarray}
\mathcal{M}&=&\left(\begin{array}{ccc}
\frac{\pi^{0}}{\sqrt{2}}+\frac{\eta}{\sqrt{6}}&\pi^{+}&K^{+}\\
\pi^{-}&-\frac{\pi^{0}}{\sqrt{2}}+\frac{\eta}{\sqrt{6}}&
K^{0}\\
K^- &\bar{K}^{0}&-\frac{2\eta}{\sqrt{6}}
\end{array}\right).
\end{eqnarray}

In Ref. \cite{falk}, the coupling constant $g=0.75$ was estimated
roughly within the quark model. A different set of coupling
constants can be found in Ref. \cite{behill}. With our notation,
$g=0.6$ \cite{behill}. In fact, the coupling constant $g$ was
studied using many theoretical approaches such as QCD sum rules
\cite{QSR,QSR-1,QSR-2,QSR-3}. Despite so many theoretical
estimates of the coupling constant $g$, we use the value
\begin{eqnarray}
g=0.59\pm 0.07\pm0.01
\end{eqnarray}
in this work. The above value was extracted by fitting the precise
experimental width of $D^*$ \cite{isoda}. In order to estimate the
values of the coupling constant $g_{\sigma}$, we compare the
Lagrangian with that in Ref. \cite{behill} and get
\begin{eqnarray}
g_\sigma={g_\pi\over {2\sqrt6}}
\end{eqnarray}
with $g_\pi=3.73$. Unlike the case of $Z^+(4430)$ \cite{xiangliu},
it is unnecessary to care about the phases of the coupling
constants in the present case. We will turn to this point later.

\section{The Derivation of the One Pion
and Sigma Exchange Potential}\label{sec3}

To derive the effective potential, we follow the same procedure in
Ref. \cite{xiangliu}. Firstly we derive the elastic scattering
amplitudes of both the direct process and crossed channel.
Secondly, we get the potential in the momentum space for a special
component (e.g. $J_z=0$) with the Breit approximation. Then we
average the potential in the momentum space. Finally we make
Fourier transformation to derive the potential in the coordinate
space.

\begin{figure}[htb]
\centering \scalebox{0.54}{\includegraphics{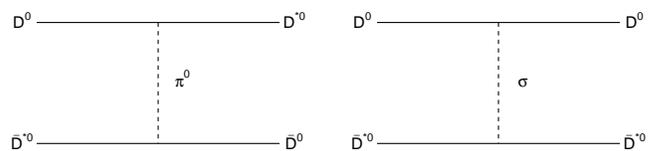}}
\caption{The scattering of $D^{0}-\bar{D}^{*0}$ by exchanging the
$\pi$ and $\sigma$ mesons. \label{haha} }
\end{figure}

In the present case, the parity and angular momentum conservation
ensures that the $\pi$ exchange occurs only in the crossed channel
while the $\sigma$ exchange only in the direct channel (see Fig.
\ref{haha}). The zeroth component of exchange meson momentum is
$q_0\approx M_{i}-M_{f}$. For the direct scattering diagram,
$M_{i,f}$ denotes the mass of $D^0$. Thus we can approximately
take $q_0=0$ and $q^2=-\mathbf{q}^2$.

However, $q_0$ could not be ignored because $M_i$ and $M_f$ denote
respectively the masses of $D^0$ and $D^{*0}$ for the crossed
diagram. $q_0=M_{D^{\ast 0}}-M_{D^0}$ is larger than pion mass
$m_\pi$ which indicates that the exchanged pion can be on-shell.
In this case, one can deal with the potential in the coordinate
space by the principal integration as in Eq. (\ref{PI}) below.

We use the following definitions in the potentials after Fourier
transformation
\begin{eqnarray}
Y_\sigma(\mathbf{r})&=&\int\frac{1}{\mathbf{q}^2+m_\sigma^2}e^{i\mathbf{q}\cdot\mathbf{r}}\frac{d\mathbf{q}}{(2\pi)^3},\\
Y_\pi(\mathbf{r})&=&\int{\cal
P}\Big[\frac{\mathbf{q}^2}{q^2-m_\pi^2}e^{i\mathbf{q}\cdot\mathbf{r}}\Big]\frac{d\mathbf{q}}{(2\pi)^3}.\label{PI}
\end{eqnarray}
Writing them explicitly, we have
\begin{eqnarray}\label{XXX}
Y_\sigma(\mathbf{r})&=&\frac{1}{4\pi r}e^{-m_\sigma r},\nonumber\\
Y_\pi(\mathbf{r})&=&-\delta(\mathbf{r})-\frac{\mu^2}{4\pi r}\cos(\mu
r),
\end{eqnarray}
where $\mu=\sqrt{q_0^2-m_\pi^2}$. Except the relative sign,
$Y_\pi(\mathbf{r})$ is similar to the expression derived in Ref.
\cite{suzuki} by using the polarization vectors
$\epsilon^{\pm1}=\frac{1}{\sqrt{2}}(0,\pm1,i,0)$ and
$\epsilon^{0}=(0,0,0,-1)$ \footnote{We have reached agreement on
the relative sign through helpful correspondence with Dr M.
Suzuki.}.

With the convention of the X(3872) flavor wave function in Eq.
(\ref{Xwave}), the potential in the study of the molecular picture
finally reads as
\begin{eqnarray}\label{Xpotential}
V(\mathbf{r})=g_\sigma^2 Y_\sigma(\mathbf{r})+\frac{g^2}{6f_\pi^2}
Y_\pi(\mathbf{r}).
\end{eqnarray}
Here the sign between one sigma exchange potential (OSEP) and OPEP
is determined by the relative sign of $|D^0 \bar{D}^{*0}\rangle$
and $|{D}^{*0}\bar{D}^0 \rangle$ in the wave function in Eq.
(\ref{Xwave}).

It's important to note that the signs in the potential are
completely fixed. The heavy quark spin-flavor symmetry ensures
that the $D$ and $\bar{D}^\ast$ mesons possess the same coupling
constants. The resulting potential in Eq. (\ref{Xpotential}) does
not change with the phases of coupling constants.

Especially, we find that $\sigma$ exchange potential is repulsive,
which differs from that in the nuclear forces. Because of this
unique feature, one just needs to study whether the one-pion
exchange can bind $D$ and $\bar{D}^\ast$ mesons to form $X(3872)$.
Only when the answer is positive, should we consider the effect
from the $\sigma$ exchange.

We note that the potential in Eq. (\ref{Xpotential}) is derived with
the implicit assumption that all the mesons are point-like
particles. Such an assumption is not fully reasonable due to the
structure effect in every interaction vertex depicted in Fig.
\ref{haha}. Thus in the following we will introduce the cutoff to
regulate the potential and further study whether it is possible to
find a loosely bound molecular state using the realistic potential.

We will modify the potential through two approaches: (1)
considering the form factor (FF) contribution; (2) smearing the
potential. Although these two approaches look different, they are
essentially the same, i.e. imposing a short-distance cutoff to
improve the singularity of the effective potential.

\subsection{Introducing form factors in the potential}

Before making a Fourier transformation, we introduce a form factor
in the interaction vertex to compensate the off-shell effects of the
exchanged mesons. The adopted FF is of the monopole type
\cite{Tornqvist,FF}
\begin{eqnarray}
F(q)=\frac{\Lambda^2-m^2}{\Lambda^2-q^2},
\end{eqnarray}
where $\Lambda\sim 1 \,\mathrm{GeV}$ denotes a phenomenological
cutoff. $m$ and $q$ are the mass and the four-momentum of the
exchanged meson respectively. As $q^2\to 0$, FF becomes a
constant. With $\Lambda\gg m$, it approaches unity. In other
words, as the distance is infinitely large, the vertex looks like
a perfect point. So the form factor is simply unity. On the other
hand, as $q^2\rightarrow \infty$, the form factor approaches to
zero. In this situation, as the distance becomes very small, the
inner structure (quark, gluon degrees of freedom) would manifest
itself and the whole picture of hadron interaction is no longer
valid.

\begin{figure}
\scalebox{0.5}{\includegraphics{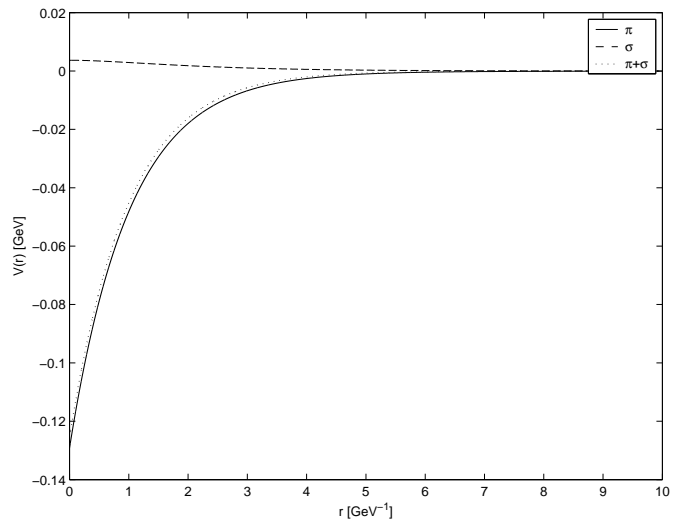}} \caption{The
regulated potentials related with $X(3872)$ in the case of FF. The
solid line corresponds to OPEP. The long-dash line comes from one
$\sigma$ exchange, and the short-dash line is the total effective
potential. $g=0.59$, $g_\sigma=0.76$ and $\Lambda=1.0$ GeV are
used. \label{potential_FF}}
\end{figure}

The explicit expressions of the modified potentials are
\begin{eqnarray}
Y_\sigma(r)&=&\frac{1}{4\pi r}(e^{-m_\sigma r}-e^{-\Lambda
r})-\frac{\eta'^2}{8\pi\Lambda}e^{-\Lambda r},\\
Y_\pi(r)&=&-\frac{\mu^2}{4\pi r}[\cos(\mu r)-e^{-\alpha
r}]-\frac{\eta^2\alpha}{8\pi}e^{-\alpha r},
\end{eqnarray}
where $\eta=\sqrt{\Lambda^2-m_\pi^2}$,
$\eta'=\sqrt{\Lambda^2-m_\sigma^2}$ and
$\alpha=\sqrt{\Lambda^2-q_0^2}$. Note we use the same $\Lambda$
for $\pi$ and $\sigma$ exchange. As an example, we have plotted
the above regulated potential in Fig. \ref{potential_FF}.

\subsection{Regulating the potential with the smearing technique}

The potential can be written as
\begin{eqnarray}
&&V(\mathbf{r})=\int
V(\mathbf{r}')\delta(\mathbf{r}-\mathbf{r}')d\mathbf{r}'\,.
\end{eqnarray}
To smear the potential, we employ the replacement
\begin{eqnarray}
\delta(\mathbf{r}-\mathbf{r}')\rightarrow
\left(\frac{\beta}{\pi}\right)^{3/2} e^{-\beta
(\mathbf{r}-\mathbf{r}')^2},
\end{eqnarray}
which was suggested by Isgur in Ref. \cite{ISGUE}. As $\beta$ goes
to infinity, the right-hand-side of the above expression becomes
the delta function. Typical values of $\sqrt{\beta}$ are
$\sqrt{\beta}\sim 1$ GeV, corresponding to the short range cutoff.
I.e., the short-distance structure is indiscriminate. On the other
hand, $\beta$ should not be very small to describe a system with
internal structure.

\begin{figure}
\centering \scalebox{0.5}{\includegraphics{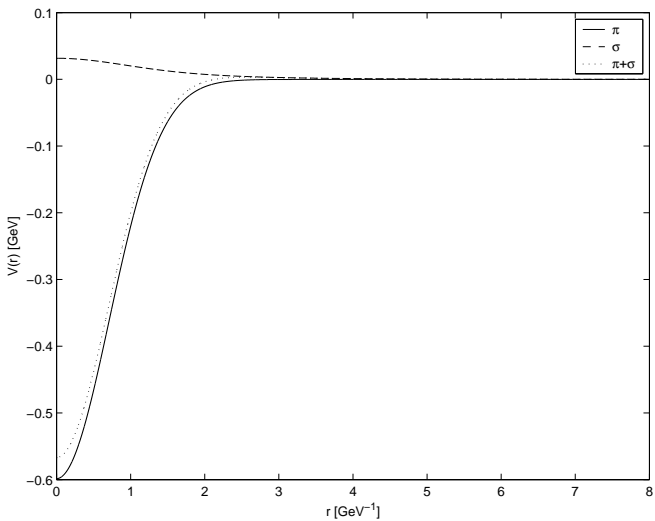}}
\caption{The regulated potentials related with $X(3872)$ in the
case of smearing. The solid line corresponds to OPEP. The
long-dash line comes from one $\sigma$ exchange, and the
short-dash line is the total effective potential. $g=0.59$,
$g_\sigma=0.76$ and $\beta=1.0$ GeV$^2$ are
used.\label{potential_S}}
\end{figure}

We obtain the smearing potential
\begin{eqnarray}
&&V(r)_{\tt smearing}\nonumber\\
&=&\frac{g_{\sigma}^2}{8\pi r}e^{-\beta
r^2}\Big[e^{\frac{(m_{\sigma}-2\beta r)^2}{4\beta}}{\tt
erfc}\Big(\frac{m_{\sigma}-2\beta
r}{2\sqrt{\beta}}\Big)\nonumber\\&&-e^{\frac{(m_{\sigma}+2\beta
r)^2}{4\beta}}{\tt erfc}\Big(\frac{m_{\sigma}+2\beta
r}{2\sqrt{\beta}}\Big)\Big]-\frac{g^2}{6f_{\pi}^2}\Big(\frac{\beta}{\pi}\Big)^{3/2}e^{-\beta{r}^2}\nonumber\\&&-\frac{g^2\mu^2
e^{-\beta r^2}}{48f_{\pi}^2\pi r}\Big[ e^{\frac{(2\beta
r-i\mu)^2}{4\beta}} {\tt erf}\Big(\frac{2\beta
r-i\mu}{2\sqrt{\beta}}\Big)+c.c.\Big].
\end{eqnarray}
Here $\mathrm{erf(x)}$ and $\mathrm{erfc(x)}$ denote the error
function and complementary error function respectively while
$c.c.$ denotes complex conjugate. An illustrative example of the
smeared potential is presented in Fig. \ref{potential_S}.

\section{Numerical Results From the One Pion Exchange Interaction Alone}
\label{sec4}

In order to find whether there is a bound state in $D\bar{D}^\ast$
system, we solve the radial Schr\"odinger equation with the help of
MATSLISE \cite{matslise}, which is a graphical MATLAB package for
the numerical solution of Sturm-Liouville and Schr\"{o}dinger
equations. A bound system has at least one negative eigenvalue.

To solve the Schr\"odinger equation, one needs the following
parameters: $m_\pi=134.98$ MeV, $m_\sigma=600$ MeV, $f_\pi=132$
MeV, $m_{D^\ast}=2006.7$ MeV, $m_{D^0} = 1864.6$ MeV \cite{PDG}.
In this section, we first consider whether the one pion exchange
interaction alone can bind $D\bar{D}^\ast$.

Now we explore at what condition $D$ and $\bar{D}^*$ can form a
bound state through one pion exchange interaction with two
approaches. Our procedure to collect the numerical values is: (1)
we fix the coupling constant $g=0.59$ and vary the cutoff
($\Lambda$ or $\beta$) from a small value until we find a solution
with a binding energy less than 5 MeV; and (2) we increase $g$ to
several larger numbers and tune the cutoff until a solution with a
binding energy less than 5 MeV is found.

\subsection{Results for the case of FF}

If the coupling constant g is fixed to be the experimental value
$g=0.59$, the possible bound state solution with a negative
eigenvalue can only be found when $\Lambda > 5.6$ GeV. The larger
the cutoff $\Lambda$ is, the closer the regulated potential is to
the delta function, hence the larger the binding energy. The binding
energy is very sensitive to $\Lambda$. This result is consistent
with the behavior that $F(q^2)\rightarrow 1$ when
$\Lambda\rightarrow\infty$. It's known that the three-dimensional
$-\delta({\mathbf r})$ function alone does not generate a bound
state. The requirement $\Lambda > 5.6$ GeV is much much larger than
the commonly used reasonable value $\sim 1.0$ GeV. In other words,
the one pion exchange potential alone does NOT bind the $D^0{\bar
D}^{\ast 0}$ pair into a molecular state with the physical values of
g and $\Lambda$! This is our first important observation.

\begin{table}[htb]
\centering
\begin{tabular}{c|cccc}\hline
&$\Lambda$ (GeV)&$E_0$ (MeV) &$r_{\rm rms}$ (fm) & $r_{\rm max}$
(fm)
\\\hline & 5.7 & -0.3 & 5.8 & 0.2
\\\raisebox{1.3ex}[0pt]{$g=0.59$}& 5.8 & -2.1 & 2.2 & 0.2
\\\hline  & 4.1 & -0.8 & 3.7 & 0.2
\\\raisebox{1.3ex}[0pt]{$g=0.7$} & 4.2 & -3.2 & 1.8 & 0.2
\\\hline  & 3.1 & -0.1 & 8.7 & 0.4
\\       $g=0.8$ & 3.2 & -1.6 & 2.6 & 0.3
\\        & 3.3 & -4.9 & 1.5 & 0.2
\\\hline  & 2.5 & -0.6 & 4.2 & 0.4
\\\raisebox{1.3ex}[0pt]{$g=0.9$} & 2.6 & -2.9 & 2.0 & 0.3
\\\hline  & 2.0 & -0.2 & 7.2 & 0.5
\\\raisebox{1.3ex}[0pt]{$g=1.0$} & 2.1 & -1.8 & 2.5 & 0.4
\\\hline
\end{tabular}
\caption{Solutions for various $g$ and $\Lambda$ in the case of FF
with OPEP. Lowest eigenvalues between -5.0 MeV and -0.1 MeV are
selected.} \label{FFpi}
\end{table}

\begin{figure}
\scalebox{0.5}{\includegraphics{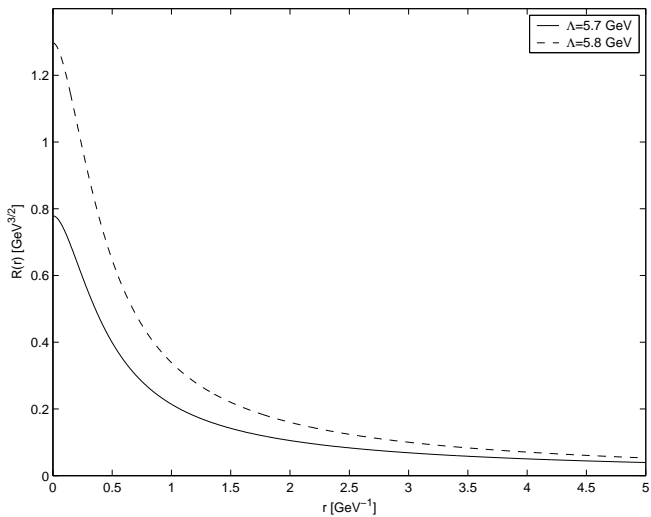} } \caption{The radial
wave functions $R(r)$ corresponding to $\Lambda=5.7$ GeV and
$\Lambda=5.8$ GeV with $g=0.59$.}\label{g59FFR}
\end{figure}

\begin{figure}[htb]
\scalebox{0.5}{\includegraphics{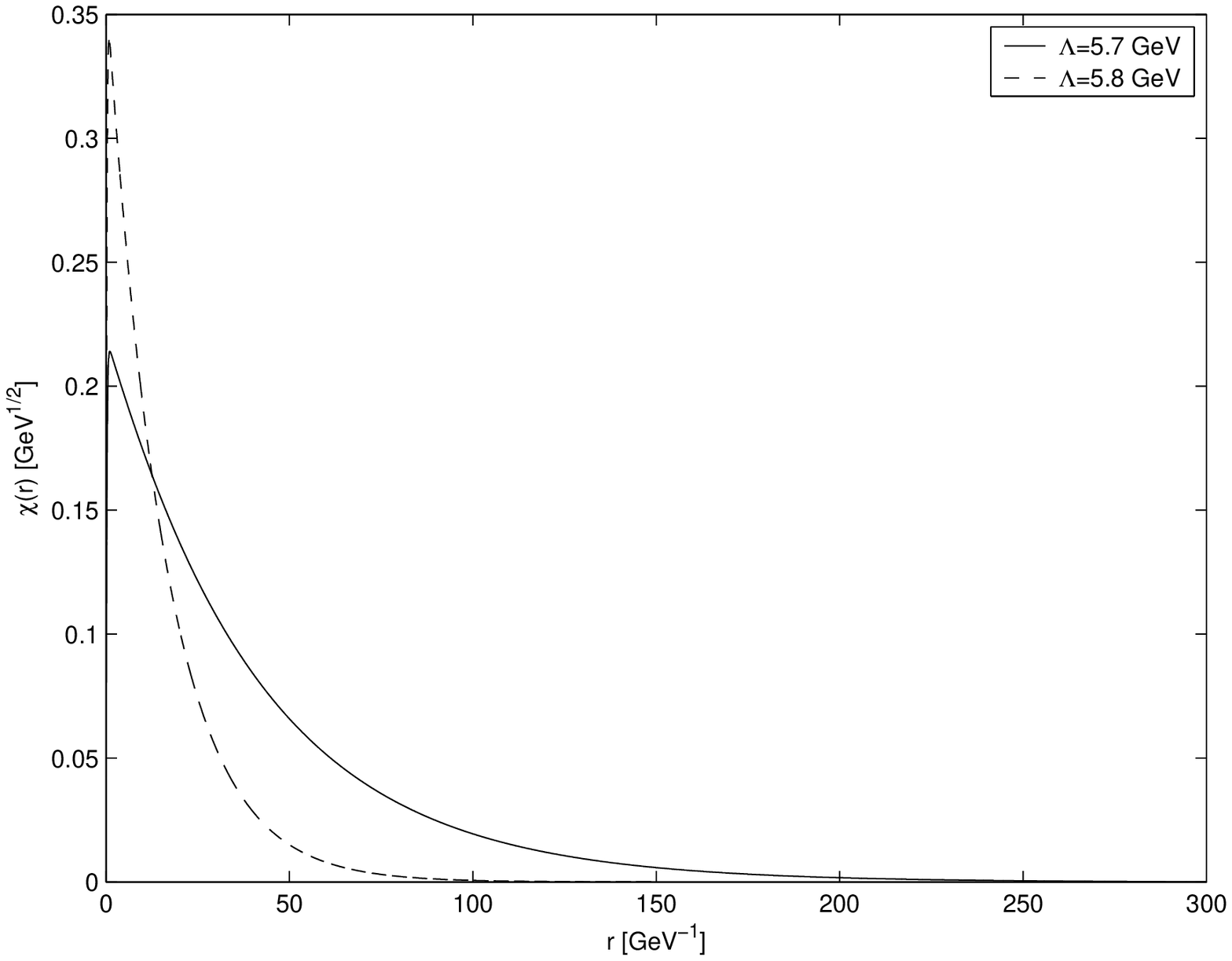} }
\scalebox{0.5}{\includegraphics{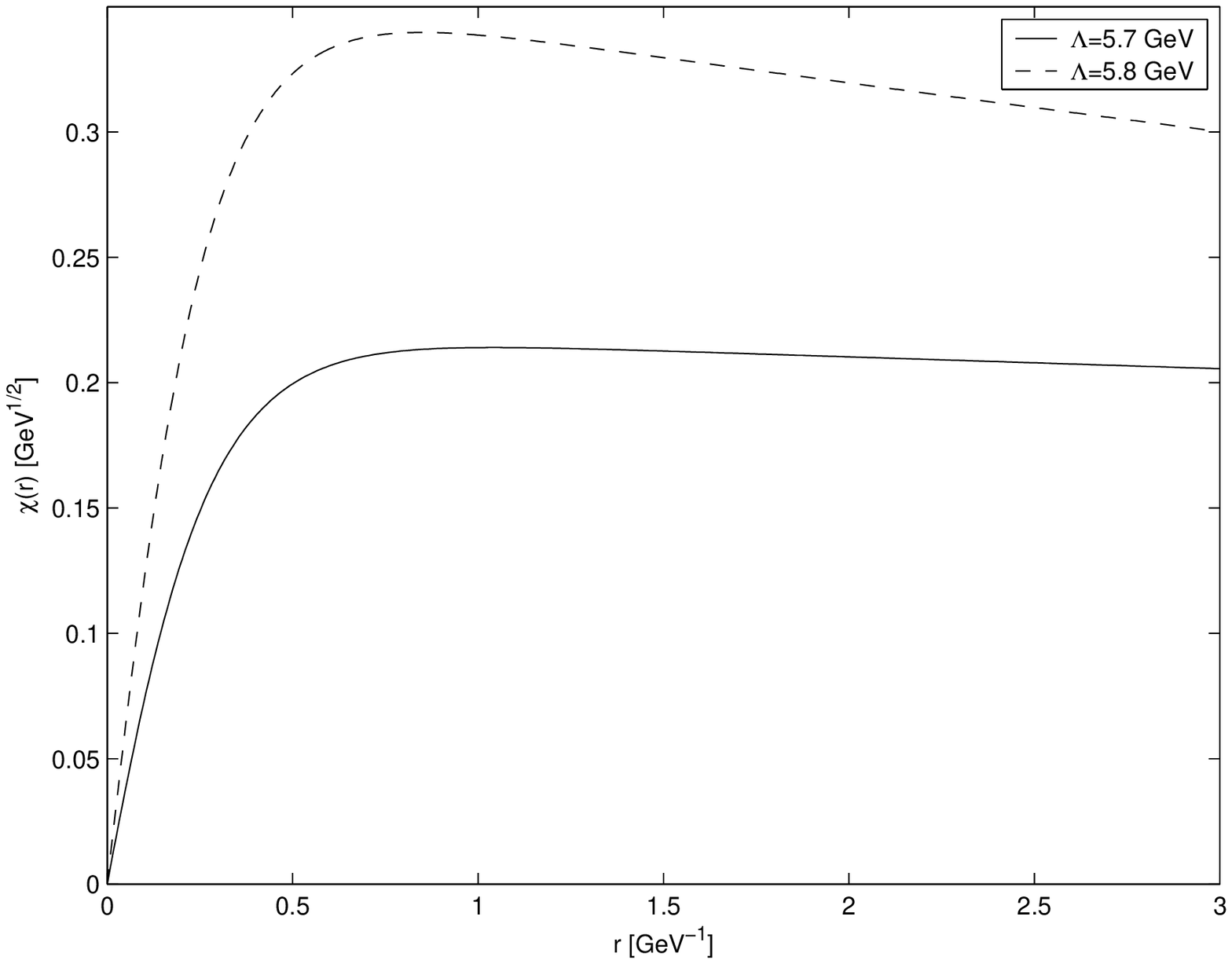}} \caption{The function
$\chi(r)=rR(r)$ corresponding to $\Lambda=5.7$ GeV and
$\Lambda=5.8$ GeV with $g=0.59$. The lower diagram shows the
behavior in short range.}\label{g59FF}
\end{figure}

We consider only the solutions with the eigenvalues between -0.1 MeV
and -5.0 MeV corresponding $\Lambda=5.7$ and $\Lambda=5.8$. To
understand the solutions more clearly, we present the numerical
results in Table \ref{FFpi}. $E_0$ is the lowest eigenvalue of the
system, $r_{\rm rms}$ is the root-mean-square radius, and $r_{\rm
max}$ is the radius corresponding to the maximum of the wave
function $\chi(r)$. In Fig. \ref{g59FFR} and \ref{g59FF}, we present
the radial wave functions $R(r)$ and $\chi(r)=rR(r)$ respectively.
According to the figures, as $\Lambda$ increases, the probability
for a bound state appearing near the origin becomes larger. The
large value of $r_{\rm rms}$ indicates this possible bound state is
very extended, which can be illustrated with the figures.

Secondly, we enlarge $g$ arbitrarily until $g=1.0$ and perform a
similar evaluation. The results are also presented in Table
\ref{FFpi}. When $g$ becomes larger, the critical point for
$\Lambda$ to generate a $D\bar{D}^\ast$ bound state becomes smaller.
With a reasonable cutoff $\Lambda \sim 1.0$ GeV, a bound state
exists only when the coupling is very strong ($g
>1.0$), which is nearly two times of the experimental value.
The wave functions corresponding to the solutions in Table
\ref{FFpi} have similar shapes with those in Figs. \ref{g59FFR}
and \ref{g59FF}.

Now we come back to discuss the partner state of X(3872). We
denote it as ${\tilde X}$. The C parity of ${\tilde X}$ is
negative.
\begin{eqnarray}
|{\tilde X}\rangle=\frac{1}{\sqrt{2}}\Big[|D^0
\bar{D}^{*0}\rangle+|{D}^{*0}\bar{D}^0 \rangle\Big].\label{XX}
\end{eqnarray}
With this convention, the signs in the OPEP are reversed while the
sigma meson exchange is still repulsive. Therefore the attractive
force is much weaker. We find that the potential is not attractive
enough to bind $D$ and $\bar{D}^\ast$ even with $g=1.0$. If we
arbitrarily use $g=5.0$ and $\Lambda=1.0$ GeV, one finds a
negative eigenvalue about -0.1 MeV. The value is not sensitive to
$\Lambda$. In this case, $r_{\rm rms}\approx19$ fm and $r_{\rm
max}$ is about 14 fm. From these values, one concludes that this
convention does not lead to a $D\bar{D}^\ast$ bound state with the
realistic coupling constant. It is not difficult to understand the
results with the potential in Eq. \ref{XXX}. The part which could
provide some attraction is $\frac{g^2\mu^2}{24\pi
f_\pi^2}\frac{\cos(\mu r)}{r}$. Since $\mu=0.044$ GeV is small, a
possible bound state exists only if $g$ is a very large number.
The consideration of FF improves mainly the behavior of the most
singular part. Thus the binding energy is insensitive to the
cutoff.

From the above analysis, we conclude that $D\bar{D}^*$ interaction
through one pion exchange is not attractive enough to form a bound
state with $g=0.59$ and $\Lambda\sim 1.0$ GeV.

\subsection{Results for the case of smearing}

In the case of the smeared potential, one fails to find a bound
state solution with negative eigenvalue for $\beta\leq 5.3 \, {\rm
GeV}^{2}$ if we fix $g=0.59$. The binding energy is very sensitive
to and increases with $\beta$. With a reasonable cutoff $\beta\sim 1
\,{\rm GeV}^2$,  there exists no loosely bound molecular state using
the realistic coupling constant $g=0.59$.

\begin{table}[htb]
\centering
\begin{tabular}{c|cccc}\hline
&$\beta$ (GeV$^2$)&$E_0$ (MeV) &$r_{\rm rms}$ (fm) & $r_{\rm max}$
(fm)
\\\hline & 5.5 & -0.3 & 5.8 & 0.2
\\       & 5.6 & -1.0 & 3.3 & 0.2
\\\raisebox{1.3ex}[0pt]{$g=0.59$}& 5.7 & -2.0 & 2.3 & 0.1
\\       & 5.8 & -3.4 & 1.8 & 0.1
\\\hline   & 2.8 & -0.3 & 5.5 & 0.2
\\       $g=0.7$ & 2.9 & -1.4 & 2.7 & 0.2
\\         & 3.0 & -3.3 & 1.8 & 0.2
\\\hline   & 1.7 & -0.9 & 3.4 & 0.3
\\\raisebox{1.3ex}[0pt]{$g=0.8$} & 1.8 & -3.1 & 1.9 & 0.2
\\\hline   & 1.1 & -1.4 & 2.8 & 0.3
\\\raisebox{1.3ex}[0pt]{$g=0.9$} & 1.2 & -4.9 & 1.5 & 0.3
\\\hline   & 0.7 & -0.5 & 4.4 & 0.4
\\\raisebox{1.3ex}[0pt]{$g=1.0$} & 0.8 & -3.9 & 1.7 & 0.3
\\\hline
\end{tabular}
\caption{Solutions for various $g$ and $\beta$ in the case of
smearing with OPEP. Lowest eigenvalues between -5.0 MeV and -0.1 MeV
are selected.} \label{smearpi}
\end{table}

When we vary $g$ from 0.59 to 1.0 and select the solutions with
$-5.0 \,{\rm MeV} < E_0 < -0.1 \,{\rm MeV}$, we obtain the results
in Table \ref{smearpi}. One gets similar conclusion as in the form
factor case. The critical point for $\beta$ to generate a bound
state is lowered as $g$ becomes larger. For example, a bound state
can be obtained with $g=0.9$ and $\beta\sim 1.0 \,{\rm GeV}^2$.
The shapes of the wave functions corresponding to these solutions
are also similar to those in Figs. \ref{g59FFR} and \ref{g59FF}.

As in the form factor case, if the flavor wave function (\ref{XX})
is used, no bound states can be found with $g=1.0$. If $g=5.0$, a
bound state exists and the eigenvalue is insensitive to the
cutoff. The numerical results are very close to those in the form
factor case, which also indicates the insensitivity of the results
to the cutoff. Therefore, it is also difficult to find a
$D\bar{D}^*$ bound state by one pion exchange interaction with the
realistic coupling constant $g=0.59$ in the smearing case.

From the above analysis within two approaches, we find that the
molecular interpretation of $X(3872)$ through one pion exchange
interaction may be problematic. The regulated OPEP may generate
bound states either with an unphysically large coupling constant
$g\ge 1.0$ or an un-reasonably large cutoff. The bound state
solution with the realistic coupling constant does not exist if the
value of the cutoff is around 1 GeV. The two approaches agrees with
each other and lead to the same conclusion. As a by-product, we
point out that our sign convention for the flavor wave function of
X(3872) is much more helpful to form a bound state than the old
convention used in the literature.

\section{Numerical Results With Both the Pion
and Sigma Meson Exchange Interaction}
\label{both}

Now we move on to include the one $\sigma$ exchange interaction.
The $\sigma$ contribution reinforces the above conclusion in the
previous section due to the repulsive nature of OSEP. We will
study carefully the variation of the numerical results and see how
much it affects the conclusion when OSEP is considered. The
procedure is similar to the OPEP case.

\subsection{Results for the case of FF}

We first take a look at the potentials plotted in Fig.
\ref{potential_FF}. The curves are obtained with $g=0.59$,
$g_\sigma=0.76$, and $\Lambda=1.0$ GeV. From this figure, one
notes that OSEP is small compared with OPEP. Thus one expects one
sigma exchange interaction has small contributions to the binding
energy. However, since a very loosely molecular state is expected,
a small variation of the potential may lead to relatively big
change of the eigenvalue.

By adding OSEP in the Schr\"odinger equation, one gets numerical
solutions listed in Table \ref{FFpi_sig}. We only use the coupling
constant $g_\sigma=0.76$ to illustrate the results. Again, we
chose the solutions with $-5.0 \,{\rm MeV}<E_0< -0.1 \,{\rm MeV}$.

\begin{table}[htb]
\centering
\begin{tabular}{c|cccc}\hline
&$\Lambda$ (GeV)&$E_0$ (MeV) &$r_{\rm rms}$ (fm) & $r_{\rm max}$
(fm)
\\\hline  & 6.0 & -1.3 & 2.8 & 0.1
\\\raisebox{1.3ex}[0pt]{$g=0.59$}& 6.1 & -4.9 & 1.5 & 0.1
\\\hline   & 4.3 & -1.1 & 3.1 & 0.2
\\\raisebox{1.3ex}[0pt]{$g=0.7$} & 4.4 & -4.5 & 1.5 & 0.2
\\\hline   & 3.3 & -0.7 & 3.8 & 0.3
\\\raisebox{1.3ex}[0pt]{$g=0.8$} & 3.4 & -3.7 & 1.7 & 0.2
\\\hline   & 2.6 & -0.4 & 5.0 & 0.3
\\\raisebox{1.3ex}[0pt]{$g=0.9$} & 2.7 & -2.8 & 2.0 & 0.3
\\\hline   & 2.1 & -0.3 & 5.9 & 0.4
\\\raisebox{1.3ex}[0pt]{$g=1.0$} & 2.2 & -2.4 & 2.2 & 0.3
\\\hline
\end{tabular}
\caption{Solutions for various $g$ and $\Lambda$ in the case of FF
with total potential. Lowest eigenvalues between -5.0 MeV and -0.1
MeV are selected. Here $g_\sigma=0.76$ is used.} \label{FFpi_sig}
\end{table}

By comparing the data in Tables \ref{FFpi} and \ref{FFpi_sig}, one
finds that many bound state solutions with negative eigenvalues
for certain pairs of $g$ and $\Lambda$ disappear after we include
the repulsive sigma meson exchange force. Only three solutions
survive with $-5.0 \,{\rm MeV} < E_0 < -0.1 \,{\rm MeV}$. But
their binding energy decreases by at least 83\%, which clearly
indicates that the sigma exchange force are numerically very
important for a loosely bound molecular state.

\subsection{Results for the case of smearing}

The smeared potentials is plotted in Fig \ref{potential_S}, where
we use $g=0.59$, $\beta=1 \,{\rm GeV}^2$ and $g_\sigma=0.76$. By
using $g_\sigma=0.76$ and selecting solutions for $E_0$ between
-5.0 MeV and -0.1 MeV, we get the results given in Table
\ref{smearpi_sig}. Comparing data in this table with those in
Table \ref{smearpi}, only two solutions (when
$g=0.9,\beta=1.2\,{\rm GeV}^2$ and $g=1.0,\beta=0.8\,{\rm GeV}^2$)
still satisfy our requirement. The binding energy decreases by at
least 74\%.

\begin{table}[htb]
\centering
\begin{tabular}{c|cccc}\hline
&$\beta$ (GeV$^2$)&$E_0$ (MeV) &$r_{\rm rms}$ (fm) & $r_{\rm max}$
(fm)
\\\hline  & 6.0 & -0.1 & 8.6 & 0.1
\\        & 6.1 & -0.8 & 3.7 & 0.1
\\\raisebox{1.3ex}[0pt]{$g=0.59$}& 6.2 & -1.9 & 2.4 & 0.1
\\        & 6.3 & -3.5 & 1.7 & 0.1
\\\hline   & 3.1 & -0.3 & 5.5 & 0.2
\\       $g=0.7$ & 3.2 & -1.6 & 2.6 & 0.2
\\         & 3.3 & -3.8 & 1.7 & 0.2
\\\hline   & 1.9 & -1.2 & 3.0 & 0.2
\\\raisebox{1.3ex}[0pt]{$g=0.8$} & 2.0 & -3.9 & 1.7 & 0.2
\\\hline   & 1.2 & -1.0 & 3.2 & 0.3
\\\raisebox{1.3ex}[0pt]{$g=0.9$} & 1.3 & -4.4 & 1.6 & 0.3
\\\hline $g=1.0$ & 0.8 & -1.0 & 3.3 & 0.3
\\\hline
\end{tabular}
\caption{Solutions for various $g$ and $\beta$ in the case of
smearing with total potential. Lowest eigenvalues between -5.0 MeV
and -0.1 MeV are selected. Here $g_\sigma=0.76$ is used.}
\label{smearpi_sig}
\end{table}

\section{Numerical results for $B\bar{B}^\ast$ system}\label{bb}

Finally we apply the formalism to $B\bar{B}^\ast$ system.
\begin{eqnarray}
|X_B\rangle=\frac{1}{\sqrt{2}}\Big[|B^+ B^{*-}\rangle-|{B}^{*+}B^-
\rangle\Big].
\end{eqnarray}
Because of the heavier masses of the B mesons, the kinematic term
has relative small contribution. The possibility of forming a
bound state is larger than that in the $D\bar{D}^\ast$ system.
OSEP remains the same. But the expression of the OPEP is different
now because $q_B^0=m_{B^\ast}-m_B < m_\pi$. Therefore the
potential can be strictly derived and does not have an imaginary
part. Now we have
\begin{eqnarray}
Y_\pi({\mathbf r})=-\delta({\mathbf r})+\frac{\mu_B^2}{4\pi
r}e^{-{\mu_B} r},
\end{eqnarray}
where $\mu_B=\sqrt{m_\pi^2-(q_B^0)^2}$.

If a form factor is introduced before the Fourier transformation,
this function becomes
\begin{eqnarray}
Y_\pi(r)=\frac{\mu_B^2}{4\pi r}[e^{-{\mu_B} r}-e^{-\alpha_B
r}]-\frac{\eta^2\alpha_B}{8\pi}e^{-\alpha_B r},
\end{eqnarray}
where $\alpha_B=\sqrt{\Lambda^2-(q^0_B)^2}$ and
$\eta=\sqrt{\Lambda^2-m_\pi^2}$.

\begin{figure}
\scalebox{0.5}{\includegraphics{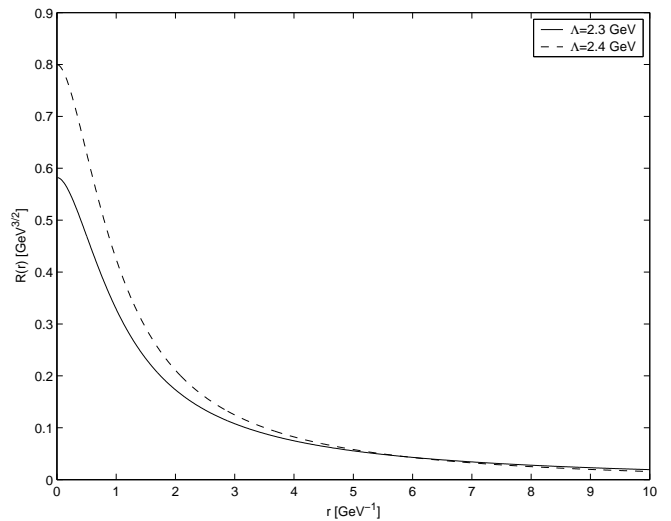} } \caption{The radial
wave functions $R(r)$ corresponding to $\Lambda=2.3$ GeV and
$\Lambda=2.4$ GeV with $g=0.59$ for the $B{\bar B}^{\ast }$
system.}\label{BR(r)}
\end{figure}

\begin{figure}[htb]
\scalebox{0.5}{\includegraphics{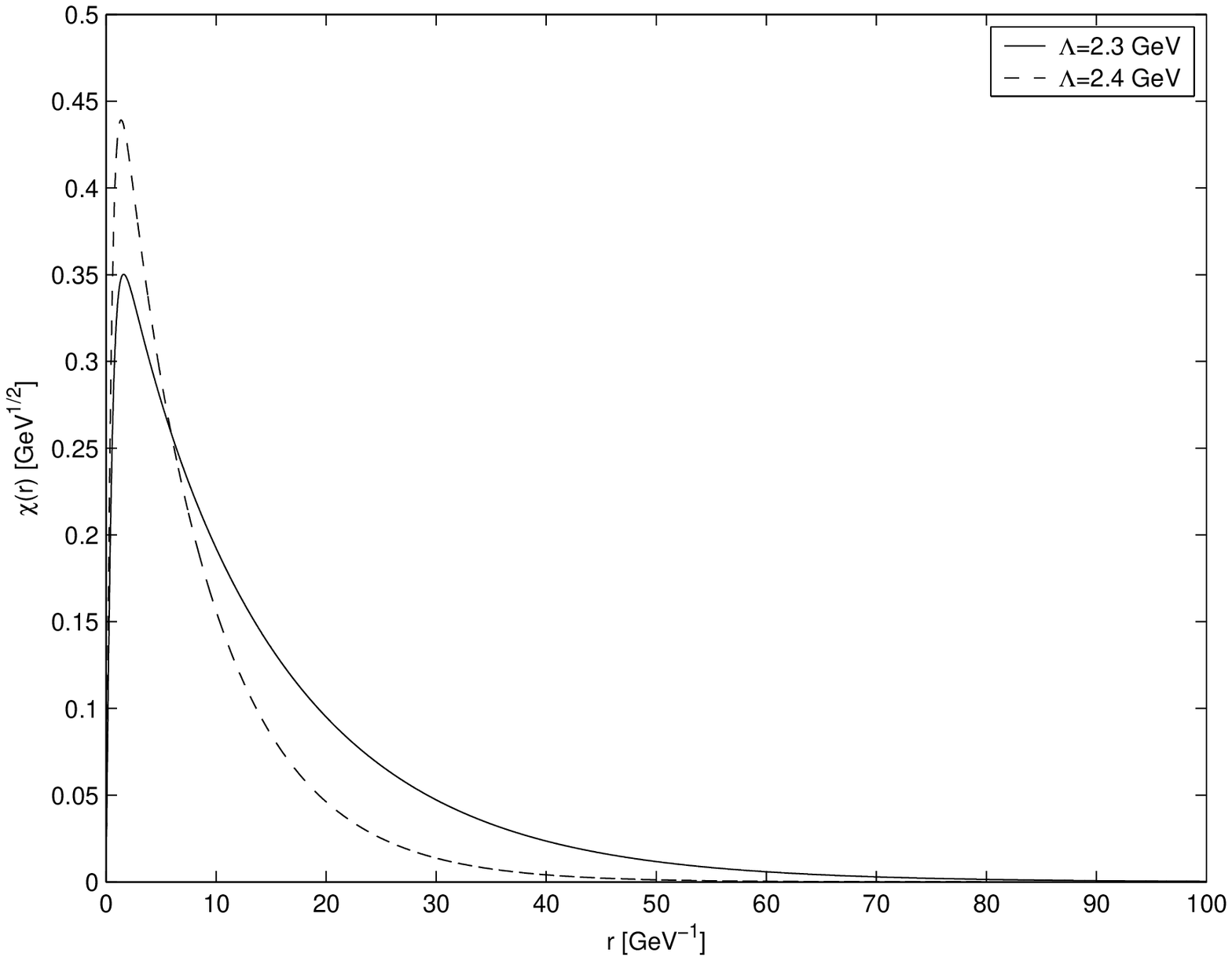} }
\scalebox{0.5}{\includegraphics{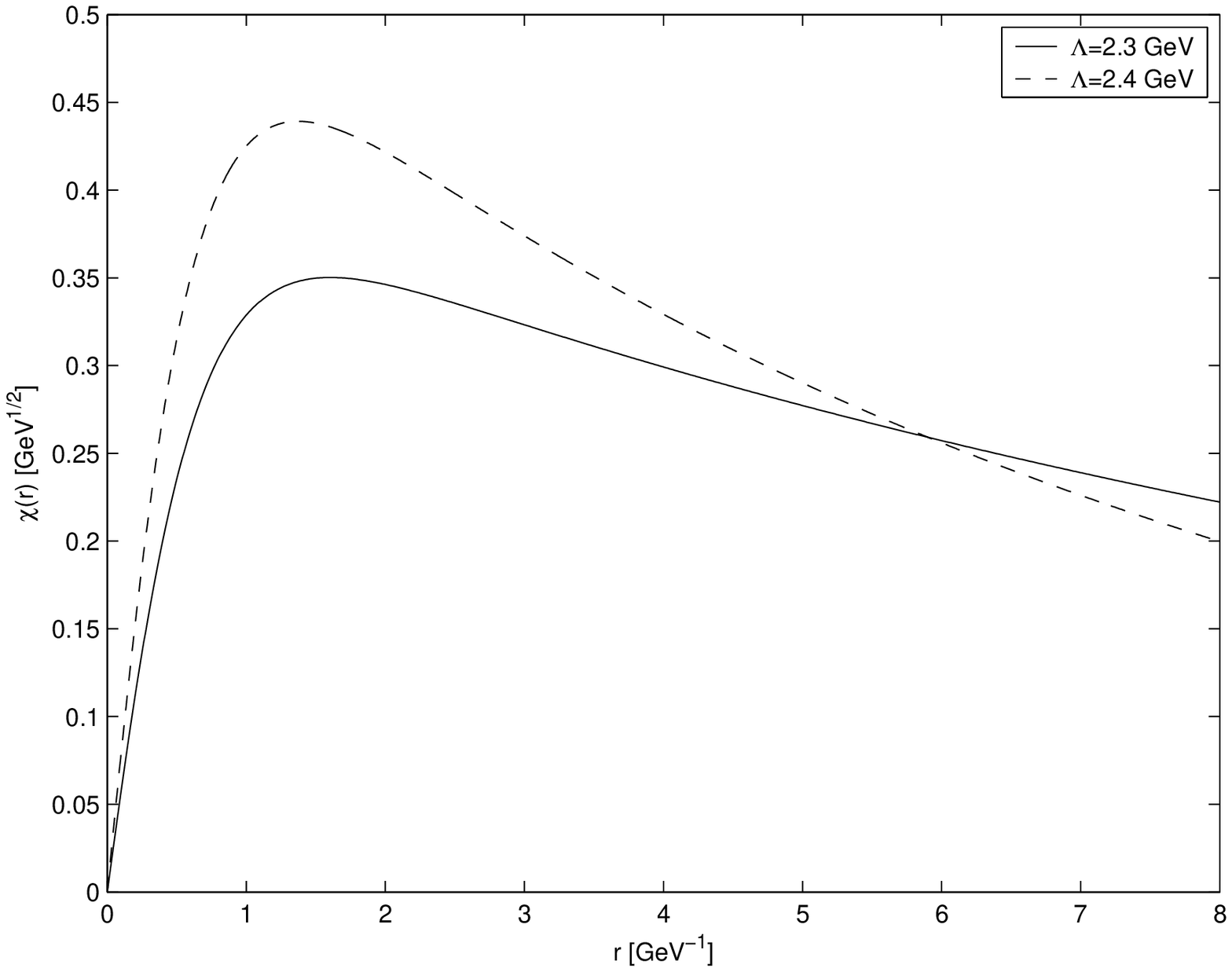}} \caption{The function
$\chi(r)=rR(r)$ corresponding to $\Lambda=2.3$ GeV and
$\Lambda=2.4$ GeV with $g=0.59$ for the $B{\bar B}^{\ast }$
system. The lower diagram shows the behavior in short
range.}\label{BChi(r)}
\end{figure}

If the smearing technique is applied, this function is regulated
as
\begin{eqnarray}
Y_\pi(r)&=&-\left(\frac{\beta}{\pi}\right)^{3/2}e^{-\beta
r^2}+\frac{\mu_B^2}{8\pi r}e^{-\beta
r^2}\nonumber\\
&&\times\left[e^{\frac{(\mu_B-2\beta r)^2}{4\beta}} {\rm
erfc}\left(\frac{\mu_B-2\beta
r}{2\sqrt{\beta}}\right)\right.\nonumber\\
&&\left.-e^{\frac{(\mu_B+2\beta r)^2}{4\beta}} {\rm
erfc}\left(\frac{\mu_B+2\beta r}{2\sqrt{\beta}}\right)\right].
\end{eqnarray}

\begin{table}[htb]
\centering
\begin{tabular}{c|cccc}\hline
&$\Lambda$ (GeV)&$E_0$ (MeV) &$r_{\rm rms}$ (fm) & $r_{\rm max}$
(fm)
\\\hline  & 2.3 & -0.9 & 2.1 & 0.3
\\\raisebox{1.3ex}[0pt]{$g=0.59$}& 2.4 & -2.8 & 1.2 & 0.3
\\\hline   & 1.7 & -0.8 & 2.3 & 0.4
\\\raisebox{1.3ex}[0pt]{$g=0.7$} & 1.8 & -2.7 & 1.3 & 0.3
\\\hline   & 1.3 & -0.1 & 5.1 & 0.6
\\       $g=0.8$ & 1.4 & -1.4 & 1.7 & 0.4
\\         & 1.5 & -4.2 & 1.1 & 0.4
\\\hline   & 1.1 & -0.4 & 3.2 & 0.6
\\\raisebox{1.3ex}[0pt]{$g=0.9$} & 1.2 & -2.3 & 1.4 & 0.4
\\\hline   & 1.0 & -1.5 & 1.7 & 0.5
\\\raisebox{1.3ex}[0pt]{$g=1.0$} & 1.1 & -5.0 & 1.0 & 0.4
\\\hline
\end{tabular}
\caption{Solutions for various $g$ and $\Lambda$ in the case of FF
for the $B\bar{B}^\ast$ system with OPEP. The lowest eigenvalues
between -5.0 MeV and -0.1 MeV are selected.} \label{XB_FFpi}
\end{table}

When performing numerical evaluations, $m_{B^\ast}=5325$ MeV and
$m_B=5279$ MeV \cite{PDG}. For the coupling constants, we use the
values in the heavy quark limit which are the same as in the
$D\bar{D}^\ast$ case. With the same procedure as before, we obtain
solutions in various cases. Results from the one pion exchange
interaction for the case of FF (smearing) are presented in Table
\ref{XB_FFpi} (\ref{XB_smearpi}). After considering effects from
the one sigma exchange interaction, the results corresponding to
the case of FF (smearing) are collected in Table \ref{XB_FFpi_sig}
(\ref{XB_smearpi_sig}). For comparison, we also present the radial
wave function R(r) and $\chi(r)$ in Figs. \ref{BR(r)} and
\ref{BChi(r)}. From these tables, it's very interesting to note
that there probably exists a loosely bound S-wave $B\bar{B}^\ast$
molecular state. Once produced, such a molecular state would be
rather stable since its dominant decay mode is the radiative decay
through $B^\ast\to B \gamma$.

\begin{table}[htb]
\centering
\begin{tabular}{c|cccc}\hline
&$\beta$ (GeV$^2$)&$E_0$ (MeV) &$r_{\rm rms}$ (fm) & $r_{\rm max}$
(fm)
\\\hline  & 0.9 & -0.9 & 2.0 & 0.3
\\\raisebox{1.3ex}[0pt]{$g=0.59$}& 1.0 & -3.2 & 1.2 & 0.3
\\\hline   & 0.5 & -1.0 & 2.0 & 0.4
\\\raisebox{1.3ex}[0pt]{$g=0.7$} & 0.6 & -4.7 & 1.0 & 0.3
\\\hline   & 0.3 & -0.5 & 2.9 & 0.5
\\\raisebox{1.3ex}[0pt]{$g=0.8$} & 0.4 & -5.0 & 1.0 & 0.4
\\\hline $g=0.9$ & 0.2 & -0.4 & 3.3 & 0.6
\\\hline $g=1.0$ & 0.2 & -4.1 & 1.1 & 0.5
\\\hline
\end{tabular}
\caption{Solutions for various $g$ and $\beta$ in the case of
smearing for the $B\bar{B}^\ast$ system with OPEP. The lowest
eigenvalues between -5.0 MeV and -0.1 MeV are selected.}
\label{XB_smearpi}
\end{table}

\begin{table}[htb]
\centering
\begin{tabular}{c|cccc}\hline
&$\Lambda$ (GeV)&$E_0$ (MeV) &$r_{\rm rms}$ (fm) & $r_{\rm max}$
(fm)
\\\hline  & 2.5 & -0.5 & 2.7 & 0.3
\\\raisebox{1.3ex}[0pt]{$g=0.59$}& 2.6 & -2.5 & 1.2 & 0.2
\\\hline   & 1.8 & -0.3 & 3.8 & 0.4
\\\raisebox{1.3ex}[0pt]{$g=0.7$} & 1.9 & -1.9 & 1.5 & 0.3
\\\hline   & 1.4 & -0.2 & 4.4 & 0.5
\\       $g=0.8$ & 1.5 & -1.7 & 1.6 & 0.4
\\         & 1.6 & -4.9 & 1.0 & 0.3
\\\hline   & 1.2 & -1.1 & 1.9 & 0.5
\\\raisebox{1.3ex}[0pt]{$g=0.9$} & 1.3 & -3.9 & 1.1 & 0.4
\\\hline   & 1.0 & -0.9 & 2.1 & 0.5
\\\raisebox{1.3ex}[0pt]{$g=1.0$} & 1.1 & -3.7 & 1.2 & 0.4
\\\hline
\end{tabular}
\caption{Solutions for various $g$ and $\Lambda$ in the case of FF
for the $B\bar{B}^\ast$ system with the total potential. The
lowest eigenvalues between -5.0 MeV and -0.1 MeV are selected.
Here $g_\sigma=0.76$ is used.} \label{XB_FFpi_sig}
\end{table}

\begin{table}[htb]
\centering
\begin{tabular}{c|cccc}\hline
&$\beta$ (GeV$^2$)&$E_0$ (MeV) &$r_{\rm rms}$ (fm) & $r_{\rm max}$
(fm)
\\\hline  & 1.1 & -0.4 & 2.9 & 0.3
\\\raisebox{1.3ex}[0pt]{$g=0.59$}& 1.2 & -2.6 & 1.2 & 0.2
\\\hline   & 0.6 & -0.6 & 2.6 & 0.3
\\\raisebox{1.3ex}[0pt]{$g=0.7$} & 0.7 & -4.1 & 1.0 & 0.3
\\\hline $g=0.8$ & 0.4 & -1.4 & 1.7 & 0.4
\\\hline $g=0.9$ & 0.3 & -3.1 & 1.2 & 0.4
\\\hline $g=1.0$ & 0.2 & -1.8 & 1.6 & 0.5
\\\hline
\end{tabular}
\caption{Solutions for various $g$ and $\beta$ in the case of
smearing for the $B\bar{B}^\ast$ system with total potential. The
lowest eigenvalues between -5.0 MeV and -0.1 MeV are selected.
Here $g_\sigma=0.76$ is used.} \label{XB_smearpi_sig}
\end{table}

\section{Summary and discussions}\label{summary}

In this work we have studied whether $X(3872)$ is an S-wave
$D\bar{D}^\ast$ molecule state bound by the one pion and one sigma
exchange interactions. We choose to work at the hadronic level and
employ the effective Lagrangian incorporating both the heavy quark
symmetry and chiral symmetry. We find the $\sigma$ meson exchange
potential is repulsive and numerically important for a loosely
bound system.

Considering the internal structure and finite size of the hadrons,
we have regulated the singular $\delta$ function in the potential
using both the form factor and smearing technique. After solving
the radial Schr\"{o}dinger equation with regulated potentials, we
find that there does NOT exist a $D^0\bar{D}^{\ast0}$
($D^{\ast0}\bar{D}^{0}$) molecular state if we use the
experimental value for the $DD^\ast\pi$ coupling constant and a
reasonable value around 1 GeV for the cutoff ($\Lambda$ or
$\sqrt{\beta}$). The two approaches lead to the same conclusion.
Bound state solutions with negative eigenvalues for the
$D\bar{D}^\ast$ system exist only with either a very large
coupling constant (two times of experimental value) or a large
cutoff ($\Lambda \sim 6$ GeV or $\beta \sim 6$ GeV$^2$).

Because B mesons are much heavier, hence their kinetic energy
decreases which is helpful to the formation of the shallow $B\bar
B$ bound state. In fact, our analysis indicates that there
probably exists a loosely bound S-wave $B\bar{B}^\ast$ molecular
state. Once produced, such a molecular state would be rather
stable since its dominant decay mode is the radiative decay
through $B^\ast\to B \gamma$. Experimental search of these states
will be very interesting.

In short summary, we have performed a dynamical calculation of the
$D^0\bar{D}^{\ast0}$ system in the mature meson exchange
framework. Our analysis disfavors the interpretation of X(3872) as
a loosely bound molecular state if we use the experimental
coupling constant and a reasonable cutoff around 1 GeV, which is
the typical hadronic scale. Clearly more theoretical and
experimental efforts are require to understand the underlying
structure of the charming and mysterious X(3872) state. Maybe one
need consider some more exotic schemes like the admixture of a
$c\bar{c}$ charmonium and a $D{\bar D}^\ast$ molecular state.
Coupled channel effects will help further lower the energy of the
system.

\vfill
\section*{Acknowledgments}

We thank Professor K. T. Chao, Professor E. Braaten, Professor M.
Suzuki, and Professor Z. Y. Zhang for useful discussions. This
project was supported by the National Natural Science Foundation
of China under Grants 10625521, 10675008, 10705001, 10775146, the
China Postdoctoral Science foundation (20060400376, 20070420526)
and Ministry of Education of China. 
\end{document}